\documentstyle[aps,preprint,prb]{revtex}
\begin{document}
\draft

\title{Anti-phase locking in a two-dimensional Josephson junction array}

\author{M. Basler\thanks{pmb@rz.uni-jena.de},
W. Krech\thanks{owk@rz.uni-jena.de} and
K. Yu. Platov\thanks{okp@rz.uni-jena.de}
}
\address{
\sl Friedrich-Schiller-Universit\"at Jena\\
\sl Institut f\"ur Festk\"orperphysik\\
\sl D-07743 Jena\\
\sl Max-Wien-Platz 1}
\date{March 6, 1996}

\maketitle
\begin{abstract}
We consider theoretically phase locking in a simple two-dimensional 
Josephson junction array consisting of two loops coupled via a joint line 
transverse to the bias current. Ring inductances are supposed to be small, 
and special emphasis is taken on the influence of external flux. Is is 
shown, that in the stable oscillation regime both cells oscillate with a 
phase shift equal to $\pi$ (i.e. anti-phase). This result may explain 
the low radiation output obtained so far in two-dimensional Josephson 
junction arrays experimentally.
\end{abstract}
\vskip2ex

\centerline{Submitted to {\it Appl. Phys. Lett.}}
\vskip2ex


Josephson Junction arrays are currently under consideration as tunable 
microwave radiation sources. \cite{lukens2,wiesenfeld1}. After clarifying 
the basic principles of (1D) 
arrays\cite{jain1,likharev1,lukens1,krech6,krech7}, there is a growing 
interest in two-dimensional (2D) arrays since the pioneering work by Benz 
and Booi\cite{benz1,benz2,booi1,sohn1}. According to basic 
estimates\cite{wiesenfeld1} radiation output generated by this type of 
arrays is expected to be much larger than that from 1D 
arrays. However, experimental results point just in the opposite 
direction. While linear arrays delivered output powers up to $50\mu$W, the 
maximum output power reported for two-dimensional arrays is around 100nW, 
\cite{stern1,kautz1,octavio1} and usually much smaller. While the general 
estimates referred to above are surely true, this should be caused by the 
fact, that actually very few junctions are locked in-phase.

There can be several reasons responsible for poor radiation output. 
Besides technological problems this can as well result from 
the fact, that the basic mechanisms of phase locking in 2D arrays,
despite some interesting results on several aspects
\cite{wiesenfeld1,filatrella1,kautz1,darula1}, have not yet been fully 
worked out theoretically. It is well known though, that there is no phase 
locking in unshunted 2D arrays in the absence of external flux. A 
theoretical study of the influence of flux with a 
''master-slave-mechanism'' by Filatrella and Wiesenfeld\cite{filatrella1} 
led to the conclusion that external flux can indeed lead to a certain 
phase locking; however the definite value of the phase difference could 
not be determined by their method, and stability was not considered at all.

Here, we start with a very simple model 2D array, consisting of two loops 
coupled via a line transverse to the bias current (Fig. 1). Despite its 
simplicity this model is difficult enough to show the essential features 
of larger arrays: It is truly two-dimensional with a possible external 
flux entering the loops and a inductance in the transverse line, as it is 
typical in the nowadays favored hybrid arrays. Our propositions are as 
follows: (i) Both junctions are considered to be identical. (ii) 
Self-inductance is taken into account while mutual inductances are 
neglected. (iii) Junctions are overdamped with $\beta\approx0$. (iv) There 
is no external load. (v) Instead of working within the framework of the 
widely used first harmonic approximation we exploit a phase slip 
technique which has proven successful in 1D arrays 
before\cite{jain1,krech6,krech7}. Its applicabiliy crucially depends on 
the proposition that the normalized ring inductance
\begin{equation}
l=2\pi I_CL/\Phi_0 
\end{equation}
($I_C$ critical current of the junctions, $L$ ring inductance, $\Phi_0$ 
flux quantum) is sufficiantly small ($l\ll1$).

Josephson junctions are described by the RSJ equations for the Josephson 
phases $\phi_{jk}$,
\begin{eqnarray}
\dot{\phi}_{jk}+\sin\phi_{jk}=i_{jk}\quad(\{j,k\}=\{1,2\}),
\end{eqnarray}
where the dot denotes differentiation w.r.t. the normalized time variable
\begin{equation}
s=\frac{2\mathrm{e}}{\hbar}R_NI_Ct
\end{equation}
($R_N$: junction normal resistance; all currents are normalized to $I_C$). 
Normalizing the external magnetic flux $\Phi$ according to 
\begin{equation}
\varphi=2\pi\Phi/\Phi_0
\end{equation}
we have to respect two flux quantization conditions,
\begin{eqnarray}
\phi_{12}-\phi_{11}-\varphi-l\overline{i}&=&0,\\
\phi_{22}-\phi_{21}-\varphi+l\overline{i}&=&0.
\end{eqnarray}
In the following the transverse current playing a crucial role in the 
coupling will be denoted by $\overline{i}$ (cf. Fig. 1). In strong 
coupling problems of this type it has proven useful to introduce sum and 
difference variables according to\cite{basler1,basler3}
\begin{eqnarray}
\label{deltasigma}
\Sigma_{k}=\frac{1}{2}(\phi_{k2}+\phi_{k1}),\\
\Delta_{k}=\frac{1}{2}(\phi_{k2}-\phi_{k1}).
\end{eqnarray}
In addition, we introduce the circular currents
\begin{equation}
i_k^{\circ}=(i_{k2}-i_{k1})/2.
\end{equation}
With the help of these variables the problem can be reformulated as
\begin{eqnarray}
\dot{\Sigma}_k+\sin\Sigma_k\cos\Delta_k=i_0,\label{sigma0}\\
\dot{\Delta}_k+\sin\Delta_k\cos\Sigma_k=i_k^{\circ},\label{delta0}\\
\Delta_1+\Delta_2-\varphi=0,\label{sumquantis}\\
\Delta_1-\Delta_2-l(i_2^{\circ}-i_1^{\circ})=0.\label{diffquantis}
\end{eqnarray}
This indicates, that the voltage sums of both loops are driven by the bias 
current $i_0>1$, while the circular currents drive voltage differences.  
Further, Eq. (\ref{sumquantis}) is the flux quantization for the whole 
array, while Eq. \ref{diffquantis} shows that differences in the circular 
currents spread the flux differences of the loops. The transverse current 
$\overline{i}$ can be obtained from
\begin{equation}
\overline{i}=i_2^{\circ}-i_1^{\circ}=\frac{1}{l}(\Delta_1-\Delta_2).
\end{equation}
According to Eq. (\ref{diffquantis}) it is just the combination 
$l\overline{i}$ which causes the coupling between the cells.

The system (\ref{sigma0})-(\ref{diffquantis}) is treated perturbatively 
assuming the ring inductance $l$ to be sufficiently small. To lowest 
order, the flux quantization conditions gives (the second index indicating 
the order of evaluation)
\begin{equation}
\Delta_{k,0}=\varphi/2,\label{delta0sol}
\end{equation}
i.e., junctions within both loops oscillate exactly in-phase. The 
Josephson oscillations itself can be evaluated from (\ref{sigma0}) as
\begin{equation}
\Sigma_{i,0}=\frac{\pi}{2}+2\arctan\frac{\zeta_0}{i_0+\cos(\varphi/2)}
\tan\left(\frac{\zeta_0s-\delta_i}{2}\right),
\end{equation}
where we introduced the flux-dependent autonomous oscillation frequency
\begin{equation}
\zeta_0=\sqrt{(i_0^2-\cos^2(\varphi/2))}.
\end{equation}
This already completes the lowest order solution for our problem; Eqs. 
(\ref{delta0}) are not required for evaluating the Josephson phases within 
this order, but determine the circular currents
\begin{equation}
i_{k,0}^{\circ}=\sin(\varphi/2)\cos\Sigma_{k,0}\label{circ0}
\end{equation}
with
\begin{equation}
\cos\Sigma_{k,0}=-\frac{\zeta_0\sin(\zeta_0 s-\delta_k)}{i_0+\cos(
\varphi/2)\cos(\zeta_0 s-\delta_k)}.
\end{equation}
To summarize, in lowest order the junctions within each cell oscillate in 
phase independently of the value of the external flux, while the relative 
oscillation phase between the cells remains undetermined.

Changing to the next order $l^1$ we start again from the Josephson phase 
differences (\ref{sumquantis}) and (\ref{diffquantis}), inserting the 
lowest order result (\ref{circ0}) on the r.h.s. of Eq. 
(\ref{diffquantis}).  From the two algebraic equations arising the 
correction terms $\Delta_{k,1}$ can be easily evaluated, and the Josephson 
phase differences of the two loops up to the first order in $l$ are given by
\begin{eqnarray}
\Delta_{1}=\frac{\varphi}{2}+\frac{l}{2}\sin(\varphi/2)(\cos
\Sigma_{2,0}-\cos\Sigma_{1,0}),\\
\Delta_{2}=\frac{\varphi}{2}-\frac{l}{2}\sin(\varphi/2)(\cos
\Sigma_{2,0}-\cos\Sigma_{1,0}).
\end{eqnarray}
From this result, one can read off the transverse current
\begin{equation}
\overline{i}=\sin(\varphi/2)\left(\cos\Sigma_{2,0}-
\Sigma_{1,0}\right)\label{i1}
\end{equation}
with the basic harmonic
\begin{equation}
\overline{i}=\frac{4\zeta_0\sin(\varphi/2)}{i_0+\zeta_0}\cos\left(\zeta_0 
s-\frac{\delta_1+\delta_2}{2}\right)\sin\left(\frac{\delta_2-
\delta_1}{2}\right).\label{transharm}
\end{equation}
We point out, that although $\overline{i}$ is proportional to $1/l$ this 
factor cancels out because of $\Delta_1-\Delta_2$ being proportional to 
$l$ itself. Accordingly, the amplitude of the transverse current is the 
same independently of the inductance $l$.

The most remarkable property of this type of ''internal shunt current'' is 
its vanishing for $\varphi=0$ and growing with the external flux 
$\varphi$. One should notice, that this behavior is just opposite to that 
of an external shunt current, which usually turns out to be proportional 
to $\cos(\varphi/2)$. The absence of any transverse rf current for 
$\varphi=0$ is however obvious: In this case the array is completely 
symmetric.

For evaluating the Josephson phase sums of the cells we exploit the method 
of ''slowly varying phase'' which has proven useful in the study of phase 
locking in one-dimensional arrays before\cite{krech6,krech7,likharev1}. 
According to this method corrections are put into the 
phases $\delta_k$,
\begin{equation}
\delta_k=\delta_k(s),
\end{equation}
which are supposed to change adiabatically only (in comparison to the rf 
Josephson oscillations) in time. In addition, we will allow for the 
possibility that the joint oscillation frequency $\zeta$ be (slightly) 
different from the autonomous frequency $\zeta_0$. With these assumptions 
the voltage sums can be written as
\begin{equation}
\dot{\Sigma_k}=\frac{\zeta_0(\zeta-\dot{\delta}_k)}{i_0+\cos(\varphi/2)
\cos(\zeta 
s-\delta_k)}.\label{sigma1}
\end{equation}
Inserting (\ref{sigma1}) into (\ref{sigma0}) and neglecting higher orders 
in $l$ after some algebra we arrive at
\begin{eqnarray}
\zeta_0(\zeta-\zeta_0-\dot{\delta_1})&=&(l/4)\overline{i}(s)\sin\varphi+
(l/2)i_0\overline{i}
(s)\sin(\varphi/2)\cos(\zeta s-\delta_1),\label{short1}\\
\zeta_0(\zeta-\zeta_0-\dot{\delta_2})&=&-(l/4)\overline{i}(s)\sin\varphi-
(l/2)i_0\overline{i}
(s)\sin(\varphi/2)\cos(\zeta s-\delta_2).\label{short2}
\end{eqnarray}
Here, all the interaction terms proportional to $l$ arising on the l.h.s. 
of Eq. (\ref{sigma0}) were transferred to the r.h.s. In this way, the 
combination $l\overline{i}$ plays a similar role as a synchronizing 
alternating external or shunt current\cite{jain1,krech8,krech11}.

To proceed, we average over one oscillation period, considering $\delta_k$ 
as roughly constant over this time interval. It can be shown, that only 
the lowest harmonic (\ref{transharm}) of $\overline{i}$ contributes. 
Evaluation of the mean values results in the evolution equations
\begin{eqnarray}
\zeta_0(\zeta-\zeta_0-<\dot{\delta}_1>)&=&l\frac{\zeta_0i_0}{2(i_0+
\zeta_0)}\sin^2(\varphi/2)
\sin(<\delta_2>-<\delta_1>),\label{delta1}\\
\zeta_0(\zeta-\zeta_0-<\dot{\delta}_2>)&=&-l\frac{\zeta_0i_0}{2(i_0+
\zeta_0)}\sin^2(\varphi/2)
\sin(<\delta_2>-<\delta_1>),\label{delta2}
\end{eqnarray}
where $<\delta_k>$ denotes the one-period average over $\delta_k$. 
Subtraction gives the reduced equation for the phase difference 
$\delta=\delta_1-\delta_2$, 
\begin{equation}
<\dot{\delta}>=l\frac{i_0}{i_0+\zeta_0}\sin^2(\varphi/2)\sin<\delta>,
\end{equation}
having formally the same structure as the RSJ equation describing an 
unbiased autonomous junction. It admits two phase locking solutions,
\begin{equation}
<\delta^{\mathrm{pl}}>=0\quad\mbox{and}\quad<\delta^{\mathrm{pl}}>=\pi,
\label{sol}
\end{equation}
describing in-phase resp. anti-phase oscillations of the cells. 
Investigation of the stability leads to the Liapunov coefficient
\begin{equation}
\lambda=l\frac{i_0}{i_0+\zeta_0}\sin^2(\varphi/2)\cos<\delta^{
\mathrm{pl}}>.\label{ljapunov}
\end{equation}
As a result, only anti-phase oscillations are stable against small 
perturbations. By substituting (\ref{sol}) into (\ref{delta1}) one easily 
recovers that the oscillation frequency remains equal to that of an 
autonomous junction, i.e.
\begin{equation}
\zeta=\zeta_0.
\end{equation}

To summarize, the following picture arises: From earlier results 
\cite{basler1,basler2} we know, that the two junctions within each 
strongly coupled cell are generally (except for $\varphi\approx\pi$) 
aligned in-phase. In addition, according to (\ref{sol}) both junctions from 
cell one oscillate anti-phase relative to those from cell 2. 
Synchronization of the cells in this state is provided by the alternating 
current (\ref{i1}), flowing through the joint transverse connection. It is 
obvious, that such a state will be non-radiating. In addition, our 
findings justify earlier results on missing phase locking in the absence 
of external flux, which within our framework can be explained by the 
marginal stability observed in (\ref{ljapunov}) for $\varphi=0$. 

All results described in this paper are in complete agreement with 
corresponding numerical simulations performed in parallel. These 
simulations show, that the observed anti-phase locking is not bounded to 
the case of small inductances treated analytically here, but is a general 
feature of this type of array. If this remains true for larger arrays, 
which is under investigation now, this might well explain the low 
radiation output obtained with two-dimensional arrays up to now. In 
addition, investigations are on the way on the interplay with an external 
shunt current. 

%%%%%%%%%%%%%%%%%%%%%%%%%%%%%%%%%%%%%%%%%%%%%%%%%%%%%%%%%%%%%%%%%%%%%
\section*{Acknowledgments}%%%%%%%%%%%%%%%%%%%%%%%%%%%%%%%%%%%%%%%%%%%
%%%%%%%%%%%%%%%%%%%%%%%%%%%%%%%%%%%%%%%%%%%%%%%%%%%%%%%%%%%%%%%%%%%%%
This work was supported by a project of the Deutsche 
Forschungsgemeinschaft DFG under contract \# Kr1172/4-1. The authors would 
like to express their thanks to the DFG for financial support.
\eject

\begin{figure}
\caption{The two-dimensional Josephson junction circuit under 
investigation.}
\label{fig1}
\end{figure}

%%%%%%%%%%%%%%%%%%%%%%%%%%%%%%%%%%%%%%%%%%%%%%%%%%%%%%%%%%%%%%%%%%%
%%%%%%%%%%%%%%%%%%%References %%%%%%%%%%%%%%%%%%%%%%%%%%%%%%%%%%%%%
%%%%%%%%%%%%%%%%%%%%%%%%%%%%%%%%%%%%%%%%%%%%%%%%%%%%%%%%%%%%%%%%%%%

\bibliographystyle{prsty}
%\bibliography{arrays}

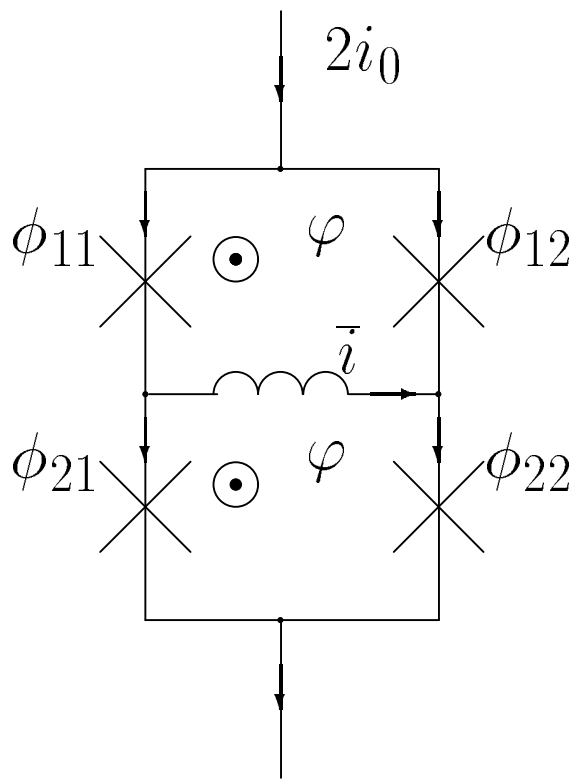
\end{document}